# Optimal Retirement of European Fossil Fuel Power Plants and the Cost of Delay

Author: Rhoden, Imke; i.rhoden@fz-juelich.de, ORCID: 0000-0001-9533-0536

Affiliations:

*Forschungszentrum Jülich GmbH, Institute of Climate and Energy Systems, Jülich Systems Analysis (ICE-2), 52425 Jülich, Germany*

*Center for Environmental Management, Resources and Energy (CURE), Faculty of Economics, Ruhr-Universität Bochum, Universitätsstr. 150, 44801 Bochum, Germany*

Abstract: Mitigation pathways require fossil power plants to retire well before the end of their technical lives, yet retirement decisions are typically derived from cost-minimizing system models rather than from the decision facing the plant's owner. This study investigates the carbon price threshold at which closure becomes more advantageous for an operator, as well as the ongoing costs associated with continued emissions. A finite-horizon Markov decision process is solved plant by plant for the European fossil fleet of 850 units, with a carbon price that moves stochastically and plants that age deterministically. The price process is constructed so that widening its range leaves the expected price unchanged, isolating the effect of uncertainty from the effect of price level. Closure thresholds fall below recent EU allowance prices for all fuels considered, and the sensitive range lies well beneath them, so further price increases are unlikely to be the binding instrument for retiring this fleet. Because fuel costs rather than emissions determine how much margin a carbon price must erode, the least carbon-intensive capacity retires first. Uncertainty raises the thresholds by about a tenth at an unchanged expected price: a credibly stable carbon price retires the same capacity as a volatile price roughly a tenth higher. Policy credibility therefore substitutes for price level.

Keywords: optimization rule, power plant phase-out, carbon pricing, option value, stranded assets, cost of delay

# 1 Introduction

Current fossil fuel power plant infrastructure could emit over 650 Gt of $CO_2$ if continued, occupying two-thirds of the remaining carbon budget until 2°C and making it difficult to reach net zero by 2050. Merely halting new construction is not sufficient, so existing plants must close before their technical lifespan ends. The energy infrastructure and electricity sector can benefit from early retirement and phased shutdown plans, as emission reductions are already cost-effective and the carbon pricing scheme is in place [1].

Accordingly, there is a gap between net-zero scenarios and actual decisions to act, as scenarios already incorporate early plant closures and pathways imply substantially shortened lifespans for energy infrastructure. In most scenarios, phasing out fossil-fuel electricity infrastructure results from cost-minimization system models or policy rather than decisions made by an actual plant operator. These scenarios indicate how much capacity should be phased out and when. However, they do not necessarily specify the conditions operators need to actually shut down a plant [2, 3]. The capital exposed by such retirement is quantified to be consistent with generation capacity at the 2°C level and at the macroeconomic level [4, 5].

In the European electricity generation sector, the carbon price set by the European Emissions Trading System (ETS 1) is the primary policy mechanism for pricing emissions, thereby making fossil-fuel electricity generation less economical. Historically, the ETS has varied considerably. From 2010 to 2025, it ranged from 3 to 100 Euro/t$CO_2$ [6]. Operators incorporate this variability into their decisions, as there is no single predetermined price due to the nature of the trading system.

Assuming the naive rule is to close a business when the operating margin turns negative, this rule contradicts actual decisions to switch off for two reasons. First, the carbon price fluctuates, so an unprofitable price today might not persist in the next decision-making period. Second, the decision to shut down or retire a plant is irreversible, as decommissioning costs are incurred, salvage value is collected, and the plant cannot be brought back once conditions improve. An operator of a fossil fuel power plant who can decide again in the next period will not close the first time the price is unprofitable, so their trigger price depends on what the operator knows about the time horizon, the decision timeframe, and how often they get to decide. A static cost-minimization calculation omits dynamic treatment of option value considerations because it treats one year in isolation, ignores that output declines with plant age, and ignores the retirement payment. Therefore, adding option value, annual degradation, and recurring decisions under uncertainty can yield different results that might suggest opposite outcomes.

To address the limitations of static cost minimization, a framework in which the operator observes a state, makes a decision, and then faces a new state is necessary. This issue can be addressed by applying a Markov decision process (MDP) to investigate 850 European fossil fuel power plants, accounting for 230.1 GW in 32 countries, in a finite-horizon model from 2025 to 2045. The operators face a stochastic carbon price and a deterministic plant age as two main state variables, and can choose between two actions. These are continuing to operate or retiring the plant, with the latter action irreversible. In this model, the carbon price process is constructed so that widening its range leaves the expected price unchanged. This isolates the effect of uncertainty from any price level, cleanly separating the option value of waiting from the price-level effect.

The model in this study uses exploratory parameters to illustrate how a decision rule can yield results that help understand plant operator strategies and the consequences of inaction. It is different from a forecast because it does not rely on estimated price dynamics but rather shows which assumptions the outcomes depend on. It is assumed that the behavior of the fossil fuel power plant fleet rests on the theory of option value. When actions are irreversible and environments are uncertain, the value of the variable determining the timing differs from when net present value is zero. Acting now removes the chance to act later under better conditions, representing an opportunity cost added to the direct cost. For investments, this means a project is triggered when its value exceeds costs. For disinvestment, owners hold onto assets beyond the zero net present value point because exit is irreversible and conditions may improve. The trigger for leaving is thus more conservative than the naive rule suggests [7, 8]. Irreversibility comes from sunk costs that cannot be recovered, making the timing of exit decisions economically significant [9].

European fossil fuel power plants face uncertainty about the carbon price, which influences the decision to close a plant permanently. The trigger is the carbon price threshold at which retirement becomes optimal. Irreversibility arises from decommissioning costs, salvage value, and the plant's absorbing state, meaning it cannot restart if the carbon price drops. Uncertainty tends to delay closure, raising the trigger price above the point where the plant's operating margin turns negative. This study models the gap between the closure threshold under known and uncertain carbon prices, assuming the same expected value. In the literature, the real options treatment of asset abandonment has been investigated for an EU coal station under uncertain carbon prices and a similar setup, but not for a whole fleet [10]. Research on natural gas power investment timing and CO2 capture and retrofitting considers only market entry, not exit, though it is a comparable counterpart [11]. A real options analysis of a single German lignite station similarly treats closure as one of several managerial choices and finds that the plant remains profitable to the end of its technical lifetime only while subsidies persist, whereas in their absence immediate shutdown is preferable [12].

The canonical economic application of this disinvestment process, or the optimal timing to retire an aging asset, is modeled using a Markov decision process framework. An agent observes the state of a machine each period and chooses between continuing to operate it and replacing it, with an optimal policy obtained by solving the model. Therefore, the structural parallel to the retirement decision is the recurring binary choice over an asset with a deteriorating state [13]. In energy, Markov decision processes have been applied predominantly to operational problems such as storage dispatch and demand-side management, and to capacity expansion planning, where the decision concerns new investment [14, 15]. Their application to the retirement of existing fossil capacity, and in particular across a whole fleet, appears to be considerably rarer [16, 17].

In contrast, the levelized cost of electricity answers whether a new plant should be built, and it includes the recovery of capital that has not yet been spent. Whether an existing plant should continue to operate is a different question, and because that capital is sunk and cannot be recovered by closing, a different answer than levelized cost of electricity is needed, one that compares the revenue and costs that closing would actually avoid, including fuel, carbon, and fixed operating costs [7].

Analyses of the drivers of EU ETS carbon prices find that a substantial share of the variation is attributable to economic activity and the growth of wind and solar electricity production, not to demand shocks [18]. A power plant operator forming expectations about future carbon costs therefore faces uncertainty rather than a predictable path, which motivates treating the price as stochastic rather than a static parameter. Investment under uncertainty further shows that policy uncertainty increases the risk premium investors require and that long-term regulatory certainty has actual value [19].

However, literature often treats asset retirement either as an output of a cost-minimizing system model or as an exposure to be measured. It does not identify the carbon price at which plant owners would choose to close them. This study therefore investigates at what carbon price it is optimal to switch off a plant and what it costs, in emissions, to keep the plant running instead.

The study proceeds with the model's methodology, carbon price process, and calibration to data in Section 2. Section 3 presents the results and offers further insights into the model's mechanism, including a robustness check and discussion. Section 4 offers policy implications and concludes.

# 2 Methodology and data

## 2.1 Model

The analysis is based on a stochastic dynamic programming model formulated as a finite-horizon Markov decision process. This framework is necessary because a static net present value calculation evaluates a single point in time and therefore cannot represent a decision that recurs. The Markov property is that the distribution of the next state depends only on the current state and the action taken, not on the sequence of states that preceded it. This makes the problem tractable as the optimal action can be determined from the current state alone, so the model does not need to track decision histories, and the number of situations to evaluate grows with the size of the state space rather than with the length of the horizon. The Markov decision process consists of a state space, a set of actions, a reward function, and transition probabilities. States describe the circumstances under which a decision is taken. Here, a state is a combination of the prevailing carbon price level and the plant's age bin, plus one absorbing state representing a retired plant. Actions are the choices available in each state, i.e., keeping the plant running or switching it off permanently. Rewards specify the payoff attached to each state-action pair. Transition probabilities give the likelihood of moving to each possible next state given the current state and the chosen action. They combine the stochastic movement of the carbon price with the deterministic aging of the plant. Solving the process yields a policy that illustrates the optimal action for every state in every period [20].

In this study, the operator does not choose once but repeatedly, and each choice is made under conditions that have since changed. Additionally, closing a plant is irreversible, and the plant cannot resume operation if conditions improve and the carbon price it faces is uncertain. This means an operator who can decide again later holds an option whose value is destroyed by acting now.

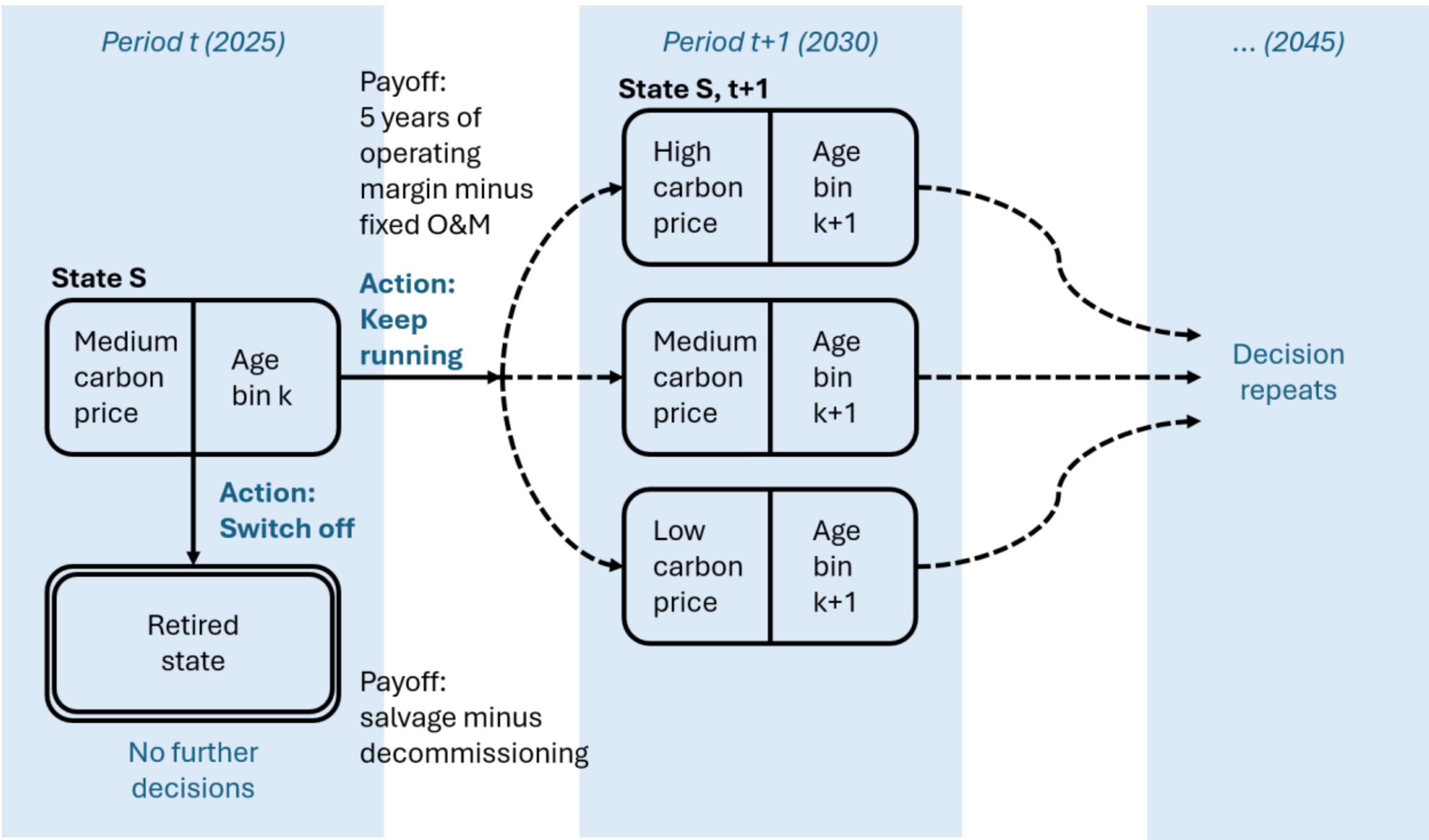

*Figure 1. Decision process for the operator of a power plant. States combine a carbon price level with an age bin; actions are to keep running or to switch off, the latter absorbing. Branching is drawn from the* medium *state only;* low *and* high *transition analogously. For transition probabilities see Table 7. Source: own illustration [20].*

For each plant in the sample, a decision process is modeled separately (cf. Figure 1). Every five years in the model's finite-horizon time frame 2025-2045, each assumed plant operator observes the carbon price and their plant's age and can choose to keep the plant running or to switch it off. Keeping the plant in operation will allow earning five years of operating profit, while switching it off is an irreversible decision that yields decommissioning costs minus salvage value, once. The carbon price moves randomly between three levels and the plant ages, and after five years, another decision option arises. The model is solved by backward induction, so that an optimal choice for every situation is generated. The resulting decision then applies to every plant for each fuel, focusing on heterogeneity of fuels, not heterogeneity within a fuel.

Emissions by plants are accumulated until 2049. By then, every plant is decommissioned at the latest, yielding the reward for switching off. This is because decommissioning is payable on every path chosen. Keeping the plant running therefore changes when the cost is paid, not whether it is paid, which is why escaping the cost by never retiring is not biasing the decision to keep a plant running. A plant's output is assumed strictly positive at every age, even though it declines with age.

All variables are listed in the appendix, Table 6. Formally, the finite-horizon MDP is solved with the Bellman recursion, where

$$W_t(s) \;=\; \max_{a\in\{1,2\}} \left\{ R_f(s,a) \;+\; \beta \sum_{s'\in\mathcal{S}} P\,(s' \mid s,a)\, W_{t+1}(s') \right\}, \qquad t = T, T-1, \dots, 1, \tag{1}$$

with terminal value $W_{T+1}$ and discount factor $\beta = (1+r)^{-\Delta}$. The optimal policy as a result of the decision rule is then the maximization

$$\pi_t(s) \;=\; \arg\max_{a\in\{1,2\}} \left\{ R_f(s,a) + \beta \sum_{s'} P\,(s' \mid s,a)\, W_{t+1}(s') \right\}. \tag{2}$$

The rewards are modeled as energy-margin-only, with production in one period following

$$q_f(k) = \kappa_f\, g_f(\alpha_k)\, H \qquad \text{MWh per MW}, \qquad H = 8760\,\Delta. \tag{3}$$

per MW of capacity, the reward function then yields

$$R_f\big((c,k),\, a\big) = \begin{cases} \underbrace{q_f(k)\left(p^e - \phi_f - \varepsilon_f\, p^c(c)\right)}_{\text{gross margin}} \;-\; \underbrace{m_f \Delta}_{\text{fixed O\&M}} & a = 1, \\ \Lambda_f \;=\; V_f - D_f & a = 2. \end{cases} \tag{4}$$

In plain terms, the reward function generates a reward for keeping a plant running consisting of MWh * (electricity - fuel - intensity * carbon), minus five years of fixed operation and maintenance (O&M). For switching off, decommissioning cost minus salvage value is paid. Both action paths are proportional to capacity, so the decision does not depend on the plant's size. The reward is further linear in the carbon price, and retirement of a plant is costly, so it is a payment, even in spite of salvage values.

Due to complexity reduction, the model is exploratory and does not contain capacity payments or scarcity rents. There is also no electricity price response to plant exit, making it a partial equilibrium perspective, as well as no ramping, commitment, repowering, or carbon-capture-and-storage retrofit. Further, no policy-forced plant closure dates are applied, and there is no plant heterogeneity within a fuel beyond age and capacity. The model is solved in R with the package MDPtoolbox [21, 22].

## 2.2 Carbon price process

The stochastic core of the MDP is the transition probability matrix. It contains the carbon prices between which the states vary, meaning that the operators facing a decision have to consider different possible carbon prices.

$$Q = \begin{pmatrix} 0.55 & 0.40 & 0.05 \\ 0.20 & 0.60 & 0.20 \\ 0.05 & 0.40 & 0.55 \end{pmatrix}, \qquad \sum_{c'} Q_{cc'} = 1 \quad \forall c \tag{5}$$

There are three possible states per period, and all revolve around a central price with a width of $\sigma = 0.5$. When $\sigma = 0$, certainty is assumed, and whenever it is not zero, uncertainty must be accounted for in decision-making. $\sigma$ is an illustrative symmetric width, based on the recent past variance of the carbon price and its observed mean price range (cf. appendix, Table 7) [6]. However, the transition matrix is a modeling choice, and it is not estimated in detail from observed price behavior.

$$p^c(\mathrm{L}) = \bar{p}\,(1-\sigma), \qquad p^c(\mathrm{M}) = \bar{p}, \qquad p^c(\mathrm{H}) = \bar{p}\,(1+\sigma) \tag{6}$$

The transition matrix spans low, medium, and high carbon prices. It is persistent, meaning that each state's most likely successor is itself. This also holds under reversal, as the probability of *high* equals the probability of *low* and the expected carbon price equals the central price (*medium*) in every period for any $\sigma$. Changing $\sigma$ therefore changes the spread of the carbon price without changing its mean, which makes any resulting change in behavior a result of uncertainty alone.

To find the optimal decision each period, the plant operator takes this matrix into account, where the expected carbon price is one of three possible outcomes. The carbon price is then factored into the calculation of the expected revenue for each period, with each state depending only on the last period's carbon price.

## 2.3 Isolation of price uncertainty

In addition to the decision rule for optimal plant closure, the analysis also allows calculation of the option value of waiting, which reflects the value of holding the option to shut off the plant at a later point in time. It can be derived because the expected carbon price has a constant mean, even though the uncertainty around it varies. Therefore, $\Omega_f > 0$ is the option value for a viable plant, following

$$\Omega_f(\sigma) \;=\; \tau_f(\sigma) - \tau_f(0), \qquad \Omega_f^{\%}(\sigma) \;=\; 100\,\frac{\tau_f(\sigma) - \tau_f(0)}{\tau_f(0)}. \tag{7}$$

The option value is measured in Euro/tCO2, as the difference between two threshold carbon prices, and in relative terms as a percentage of the threshold under certainty. It is not an emissions quantity.

The consequences of not following the resulting decision rule are measured separately in physical units. Annual emissions of a plant are given by

$$e_{it} \;=\; \frac{\mathrm{cap}_i\, \kappa_{f(i)}\, g_{f(i)}\big(\alpha_{\min(k_i+t-1,K)}\big)\cdot 8760 \cdot \varepsilon_{f(i)}}{10^6} \qquad \mathrm{MtCO_2}. \tag{8}$$

The total emissions for the time frame 2025-2049 are then reported as the difference between two scenarios, one in which all plants are kept running and the other in which the optimal decision rule estimated by the MDP is applied,

$$E^{\text{rule}} = \Delta \sum_{i=1}^{I} \sum_{t=1}^{T} \eta_{it}\; e_{it}, \tag{9}$$
$$E^{\text{open}} = \Delta \sum_{i=1}^{I} \sum_{t=1}^{T} e_{it}$$

where

$$\Delta E \;=\; E^{\text{open}} - E^{\text{rule}}. \tag{10}$$

Here, $E^{\text{open}}$ is an upper-bound scenario with all plants running, where announced closure dates are not used, and there is no dependence on the carbon price. Ω and ΔE answer different questions and are reported in different units. The first is the additional carbon price required before closure becomes optimal under uncertainty, the second is the emissions attributable to continued operation.

## 2.4 Data and calibration

The dataset includes all European power plants and is filtered to include only fossil-fuel power plants that began operations in or before 2025 [23]. Additionally, no announced plant closures before 2025 or capacity expansions after 2025 are considered. The resulting sample comprises 850 fossil-fuel power plants, totaling 230.1 GW across 32 countries. This includes 327 natural gas power plants (127 GW), 379 oil power plants (12.6 GW), 84 hard coal power plants (55 GW), and 60 lignite power plants (35.5 GW). There is a large gap between the number of plants per fuel and the electricity generated. For example, oil has the most plants, yet oil-based electricity generation is the smallest among the fuels. Additionally, the age structure varies widely by fuel, as the median commissioning year for hard coal (1972) and lignite (1976) is much earlier than for gas (2002), which is why the model's degradation factor affects the former types of power plants much more (Figure 2).

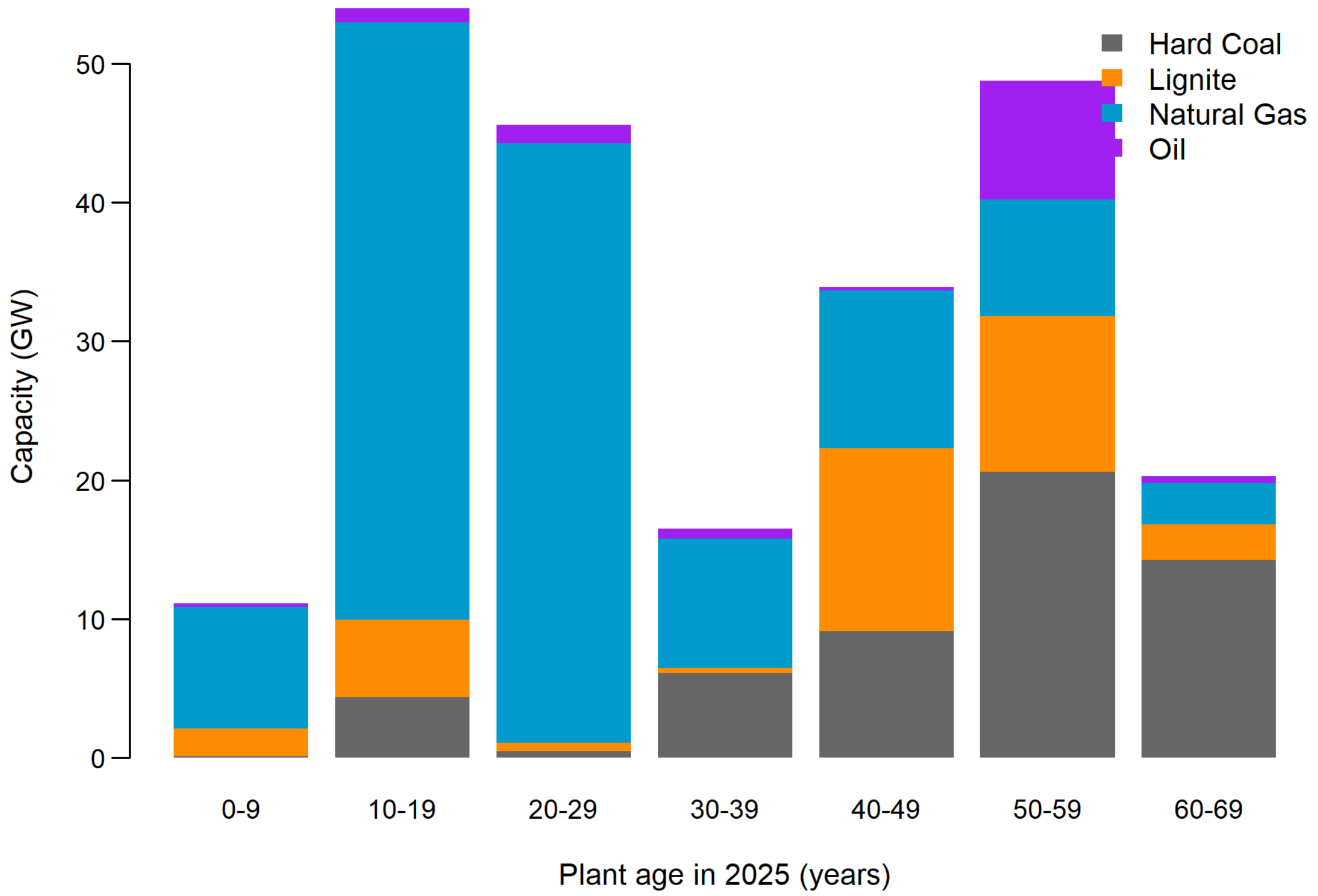

*Figure 2. Capacity of the estimation sample by fuel and 5-year-age bin in 2025. Bars are stacked, so bar height is total capacity in that bin. Source: own illustration [23].*

To represent the different fuels in the model, fuel-specific characteristics are applied according to Table 1. All values are assumptions based on the literature and are held constant to facilitate model estimation, yield exploratory results, and demonstrate the framework. Sensitivity is tested accordingly. For this analysis, the electricity price is assumed to be 80 Euro/MWh [24], flat across all countries and periods, corresponding to the 2025 average, the last full year of data included in the power plant sample. The central carbon price is assumed to be 80 Euro/tCO2, the average ETS price for 2025 and the years before [6, 25, 26]. The spread of the carbon price then corresponds to a low carbon price of 40 Euro/tCO2 and a high price of 120 Euro/tCO2. The discount rate is assumed to be 5%. The capacity factors are fleet-average values by fuel for European plants. They are not plant-specific and are estimated based on the information contained in the power plant data set and European net generation by fuel type. Decommissioning and salvage values, fixed operation and maintenance values, and annual degradation factors are sourced from the literature. Fuel costs are also fuel-specific and not plant-specific. This is mainly due to data availability issues and to keep the model estimation computationally feasible and easy to implement. The parameters therefore can only show between-fuel variation, and aside from age, no plant heterogeneity is considered. Notably, the parameters and prices chosen are post-2022, when electricity and fuel prices have skyrocketed due to geopolitical events and the war in Ukraine. Aside from the fuel parameters characterizing the fuels and the electricity and carbon price entering the reward functions, there is no feedback or nonlinearity included.

| Fuel type | Hard Coal | Lignite | Natural Gas | Oil | Sources |
|---|---|---|---|---|---|
| Carbon intensity (tCO2/MWh) | 0.867 | 1.049 | 0.358 | 1.114 | [23, 27, 28] |
| Fuel cost (Euro/MWh) | 11.6 | 2.3 | 55 | 60 | [29-32] |
| Fixed O&M (Euro/MWh/year) | 37000 | 42000 | 23000 | 30000 | [30, 33] |
| Capacity factor (0-1) | 0.19 | 0.4 | 0.24 | 0.03 | [30] |
| Decommissioning (Euro/MW) | 200000 | 200000 | 70000 | 100000 | [34, 35] |
| Salvage value (Euro/MW) | 20000 | 22000 | 15000 | 10000 | [35] |
| Annual degradation (fraction/year) | 0.01 | 0.007 | 0.004 | 0.006 | [36] |

*Table 1. Technology parameters by fuel. Capacity factor is the share of the year at full output; annual degradation compounds, so a plant of age a is reducing capacity with a factor (1 – degradation factor)^year of the initial capacity. Source: own calculation.*

# 3 Results and discussion

## 3.1 Decision rule of optimal phasing out

The standard output of an MDP is an optimal decision rule or policy. In this analysis, the decision rule indicates the optimal time to phase out a particular type of fuel. Therefore, it is representative for each fuel at an electricity price of 80 Euro/MWh.

| Fuel type | Age bin (years) | Threshold (σ = 0.5) (Euro/tCO2) |
|---|---|---|
| Hard Coal | 55+ | 46.9 |
| Lignite | 40-44 | 68.2 |
| Natural Gas | 15-19 | 44.9 |
| Oil | 55+ | 0 |

*Table 2. Switch-off thresholds by fuel for the representative plant, electricity price is at 80 Euro/MWh. Source: own calculation.*

What can be observed in Table 2 is that the order in which representative power plants are phased out does not mirror the carbon intensity of the fuels. Natural gas power plants are phased out first at a threshold carbon price of 44.9 Euro/tCO2, despite having by far the lowest carbon intensity (0.358 vs 1.049 for lignite) and being among the youngest in the sample. This is because natural gas power plants' fuel costs are 55 Euro/MWh, compared with the very low fuel costs of 2.3 Euro/MWh for lignite. This shows that the carbon price is not the only factor driving power plants to switch off.

For hard coal, the threshold is 46.9 Euro/tCO2 under an uncertain carbon price in the next period. For a representative lignite power plant, this threshold lies at 68.2 Euro/tCO2. It is the lowest central carbon price at which switching off immediately is optimal. The model returns 0 if the plant is better off closed even at a zero carbon price, which is true for the oil power plants in the sample. This means the plant is unprofitable on energy margins at any carbon price. With fixed O&M of 30,000 Euro/MW/year and a capacity factor of 0.03, fixed costs alone drive this result before any fuel is bought. However, as the oil plants in the sample are small peakers, they generate earnings from scarcity pricing and capacity payments, which the model does not include. As a result, the model's outcome is that oil power plants are not viable on energy margins alone.

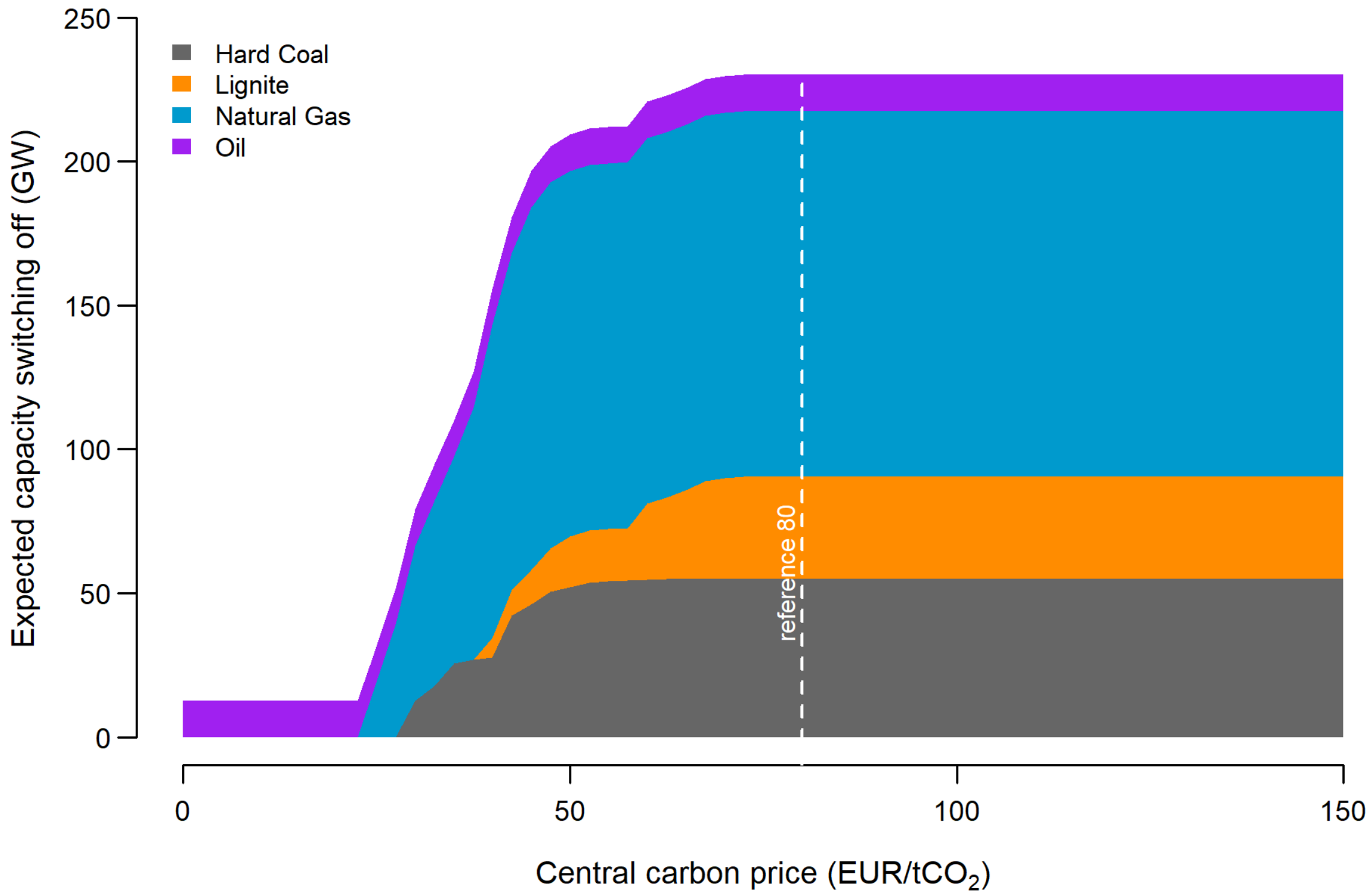

*Figure 3. Expected capacity switching off within 2025-2045, by fuel, against the central carbon price. Bands are stacked, so band thickness gives each fuel's contribution, the upper edge gives the fleet total. Capacity is weighted by the probability of switching off, with all plants starting in the* medium *carbon price state. Source: own illustration.*

Figure 3 illustrates the expected phasing out of capacity across the central carbon price. It shows the sequence in which fossil fuel power plants retire during the model's timeframe. Oil plants retire immediately at a carbon price of 0 Euro/tCO2 at the beginning of the period, in 2025. At a carbon price of 30 Euro/tCO2, 25% of the whole fleet's power generation capacity retires; 50% by a price of 37.5 Euro/tCO2; 75% by a price of 42.5 Euro/tCO2; and 99% by a price of 67.5 Euro/tCO2.

Across the different fuels, the onset of phasing out and the completion of retirements for all power plants of that fuel occur at different prices as well. Taking the first grid price of 5% and the 95% capacity of a fuel, natural gas spans carbon prices of 25 to 45 Euro/tCO2; hard coal from 30 to 52.5 Euro/tCO2; and lignite from 40 to 67.5 Euro/tCO2.

This is also shown in Figure 4. As the carbon price increases, both the total capacity and the point in time are pulled forward. At a carbon price of 30 Euro/tCO2, only 5% of capacity is retired in the first period. By 2049, 65.7% is still operating. If the carbon price is higher, e.g., 50 Euro/tCO2, 75.8% of all fossil fuel capacity is retired in the first period (2025-2029). At a carbon price of 80 Euro/tCO2, which corresponds approximately to the average ETS price, all capacity retires immediately in the first period. Any carbon price higher would similarly make all plants equally unprofitable. All four fuel thresholds are below current carbon prices, which is a property of the price. It is therefore important to focus on the sequence of the fuels being phased out, which is why the timing is reported across a range.

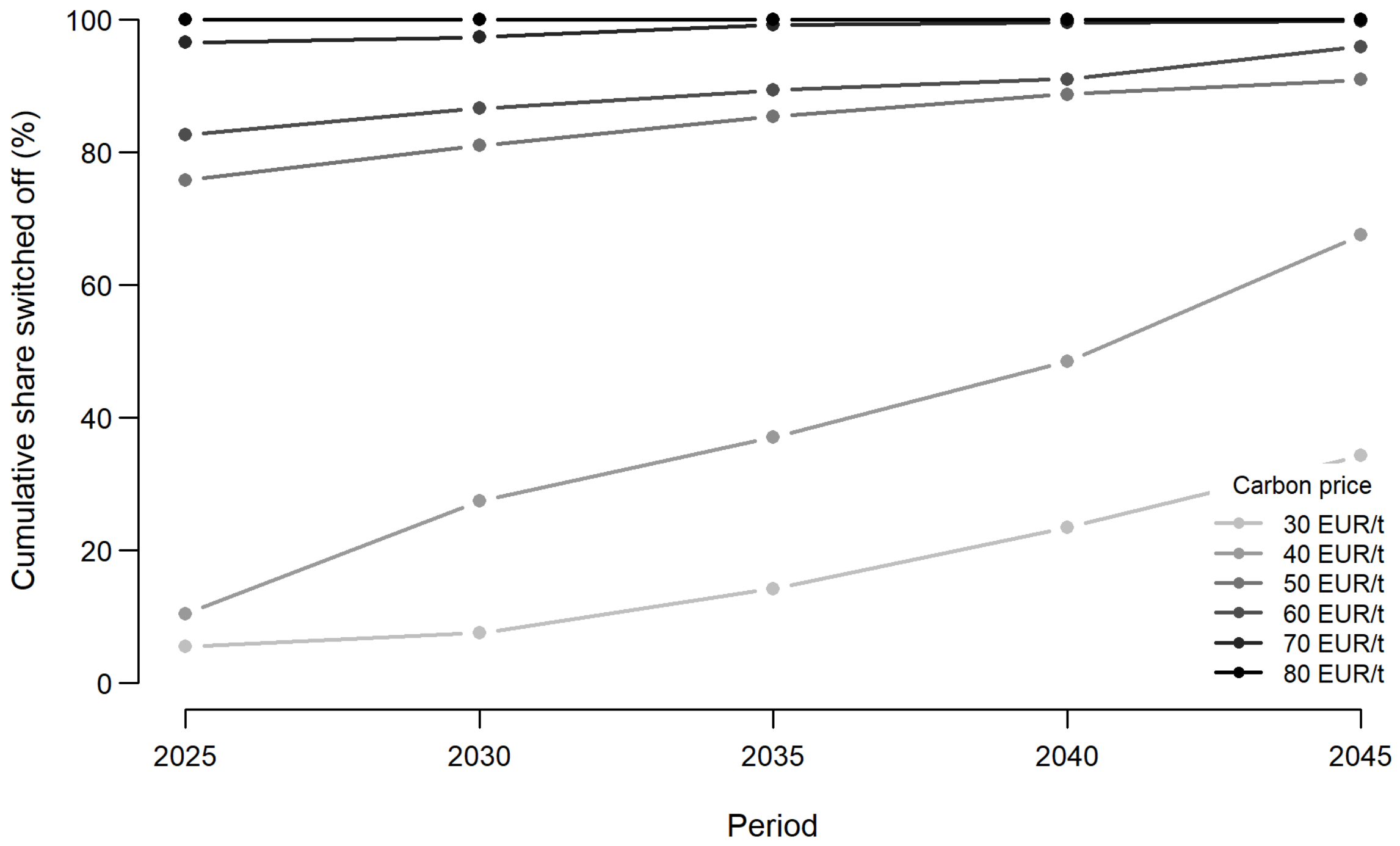

*Figure 4. Cumulative share of fleet capacity switched off by period, at six central carbon prices. Capacity that does not switch off within the horizon is the gap between each line and 100 percent. Source: own illustration.*

These thresholds can be compared with the real options literature on plant abandonment. [10] find that an EU coal station with ten years of remaining life abandons only above 162 Euro/tCO2, far above the 46.9 Euro/tCO2 obtained here. The difference is structural rather than substantive, because their plant may also close temporarily and wait. When this flexibility is removed, so that the choice is between continuous operation and permanent closure, their trigger falls to 45.6 Euro/tCO2. The model applied here offers the same two actions, so the correspondence is close. The gap between their two cases indicates that permitting temporary shutdown would raise the thresholds reported here considerably.

## 3.2 Mechanism

A core result of the MDP is that uncertainty modeled by the stochastic carbon price transition matrix leads to higher threshold prices. For hard coal, uncertainty increases the threshold price by almost 12%; for lignite and natural gas, by almost 11%. Oil is not included because it is not viable at any carbon price. Table 3 shows the effect of uncertainty. The expected carbon price is the same under both certainty ($\sigma = 0$) and uncertainty ($\sigma = 0.5$) because the transition matrix is symmetrical even when the state order is reversed (cf. appendix, Figure 8). Nothing else differs, so the rise is due only to the uncertainty caused by the possibility of the carbon price varying.

| Fuel | Certainty threshold (σ = 0) | Uncertainty threshold (σ = 0.5) | Option value change in threshold carbon price | % |
|---|---|---|---|---|
| Hard Coal | 41.9 | 46.9 | +5.0 | 11.9 |
| Lignite | 61.5 | 68.2 | +6.7 | 10.9 |
| Natural Gas | 40.5 | 44.9 | +4.4 | 10.9 |

*Table 3. Switch-off thresholds under a known and an uncertain carbon price, with the resulting option value. Both columns are computed at the same expected carbon price, Euro/tCO2. Source: own calculation.*

The effect is nearly linear in $\sigma$ and consistent across three fuels (see Figure 5). It amounts to 5.7 / 11.9 / 18.9 % for hard coal at $\sigma$ = 0.25 / 0.5 / 0.75; for lignite, it is 5.0 / 10.9 / 17.4 %, and 4.9 / 10.9 / 17.5 % for natural gas. However, the resulting change in threshold carbon prices that can be related to the option value is a direct function of $\sigma$, which is an assumed value and is kept static in the model, giving a maximum width of possible variance in the carbon prices. Still, the mechanism of increasing thresholds with increasing uncertainty and the linear behavior are valuable outcomes of the model. The threshold premium rises roughly in proportion to the width, meaning that increased uncertainty in the carbon prices will lead power plants to likely decide to phase out at higher carbon prices. The economic interpretation is that deciding to switch off a plant is irreversible. As the price might fall back, an operator who can decide again in the next period is willing to endure a carbon price about 10-11% higher than a certainty model would suggest before closing. As an example, the certainty carbon price for hard coal with $\sigma = 0$ is 41.9; with $\sigma = 0.5$, the threshold carbon price for hard coal is 46.9 Euro/tCO2. If the carbon price is 45 Euro/tCO2, in a certainty world, the plant would shut down. In an uncertain carbon price world, the plant would continue to operate even if it faced unprofitability, because the price might fall again next period.

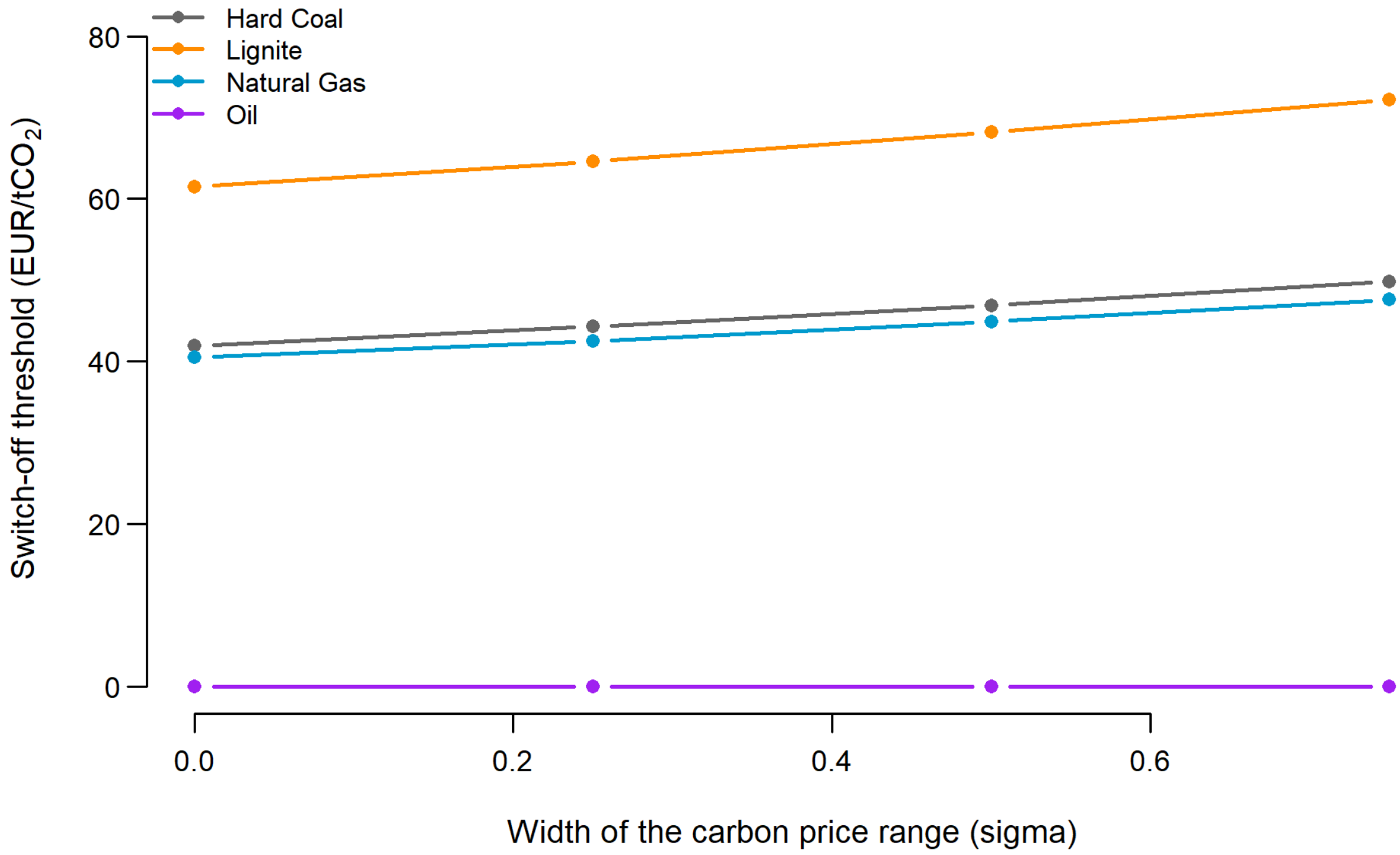

*Figure 5. Switch-off threshold against the width of the carbon price range, by fuel. The expected carbon price is identical at every width, so the vertical rise of each line is attributable to uncertainty alone. Source: own illustration.*

As a point of caution, the gap between the threshold prices at $\sigma = 0$ and $\sigma = 0.5$ should not be interpreted as a sum of identifiable effects for degradation, retirement payment, and option value. This is because these components are not separable in the model. The option value itself changes when the degradation factor is switched off due to interacting effects. The calculations are performed under different assumptions and should not be read as subsequent steps.

| Fuel | Width ($\sigma$) | Threshold known (Euro/tCO2) | Threshold uncertain ( $\sigma =$ 0.5) (Euro/tCO2) | OptionValue (Euro/tCO2) | OptionValue (%) |
|---|---|---|---|---|---|
| Hard Coal | 0.25 | 41.9 | 44.3 | 2.4 | 5.7 |
| Hard Coal | 0.5 | 41.9 | 46.9 | 5 | 11.9 |
| Hard Coal | 0.75 | 41.9 | 49.8 | 7.9 | 18.9 |
| Lignite | 0.25 | 61.5 | 64.6 | 3.1 | 5 |
| Lignite | 0.5 | 61.5 | 68.2 | 6.7 | 10.9 |
| Lignite | 0.75 | 61.5 | 72.2 | 10.7 | 17.4 |
| Natural Gas | 0.25 | 40.5 | 42.5 | 2 | 4.9 |
| Natural Gas | 0.5 | 40.5 | 44.9 | 4.4 | 10.9 |
| Natural Gas | 0.75 | 40.5 | 47.6 | 7.1 | 17.5 |
| Oil | 0.25 | 0 | 0 | 0 | |
| Oil | 0.5 | 0 | 0 | 0 | |
| Oil | 0.75 | 0 | 0 | 0 | |

*Table 4. Switch-off thresholds under a known and an uncertain carbon price, with the resulting option value. Both columns are computed at the same expected carbon price. The expected carbon price is identical in every row, so the difference between the two threshold columns is the effect of uncertainty alone. Source: own calculation.*

The direction of this effect is consistent with earlier findings. Higher allowance price volatility raises the carbon price required for abandonment, for the same reason identified here [10]. Further, policy uncertainty raises the risk premium investors require [19]. However, the magnitudes are not comparable, as neither previous specification in the literature holds the expected carbon price constant across volatility levels.

## 3.3 Value of a delayed phase-out

To obtain a quantitative estimate of the value of not phasing out under the optimal decision rule, the setup compares two pathways. One follows the decision rule, with plants switched off according to the MDP optimization, and the other is the counterfactual of keeping every plant running. The difference between them gives the value of delaying the phase-out. It is reported in emissions, in tonnes of CO2, because translating these values into a monetary equivalent requires a damage function, which is not included or generated by the model. Table 5 lists emissions from keeping the plants running by fuel, totaling 5,651 Mt CO2. Lignite accounts for 41% of these emissions while holding only 15% of the whole fleet's capacity, and natural gas accounts for 37% of emissions with 55% of total capacity. This shows that carbon intensity and the capacity factor drive this ranking, not power-generating capacity alone. Hard coal power plants account for 20% of all emissions, while oil emissions are negligible in comparison, at 0.01%. At a central carbon price of 80 Euro/tCO2, the gap in the delayed value of emissions amounts to the full 5,651 Mt CO2, because the path selected by the optimization in the MDP is zero, so everything closes in period 1 as it is not profitable.

| Fuel | KeepRunning (Mt CO2) | FollowRule (Mt CO2) | Difference (Mt CO2) |
|---|---|---|---|
| Hard Coal | 1162 | 0 | 1162 |
| Lignite | 2341 | 0 | 2341 |
| Natural Gas | 2080 | 0 | 2080 |
| Oil | 68 | 0 | 68 |

*Table 5. Cumulative emissions 2025-2049 by fuel under both paths, and their difference. The keep-running path ignores announced closure dates, so it is an upper bound on business-as-usual emissions rather than a forecast. Source: own calculation.*

Figure 6 visualizes the gap, which rises steeply through the threshold region marking different carbon prices. At 0-20 Euro/tCO2, 68 Mt of emissions accumulate; at 50 Euro/tCO2, it is already 3,679 Mt of emissions. The important range lies between 40 and 70 Euro/tCO2, because small price differences already exert considerable effects. At a price of 80 Euro/tCO2, there is no variation because all plants are already past their optimal phase-out time. Again, the 68 Mt CO2 floor at a carbon price of 0 Euro is due to oil power plants that retire for non-carbon-price reasons. Also, the upper line representing the counterfactual is flat because it does not respond to a carbon price. Furthermore, this path of keeping all plants running ignores announced closure dates, which is why any realistic business-as-usual cases in the literature are exceeded here.

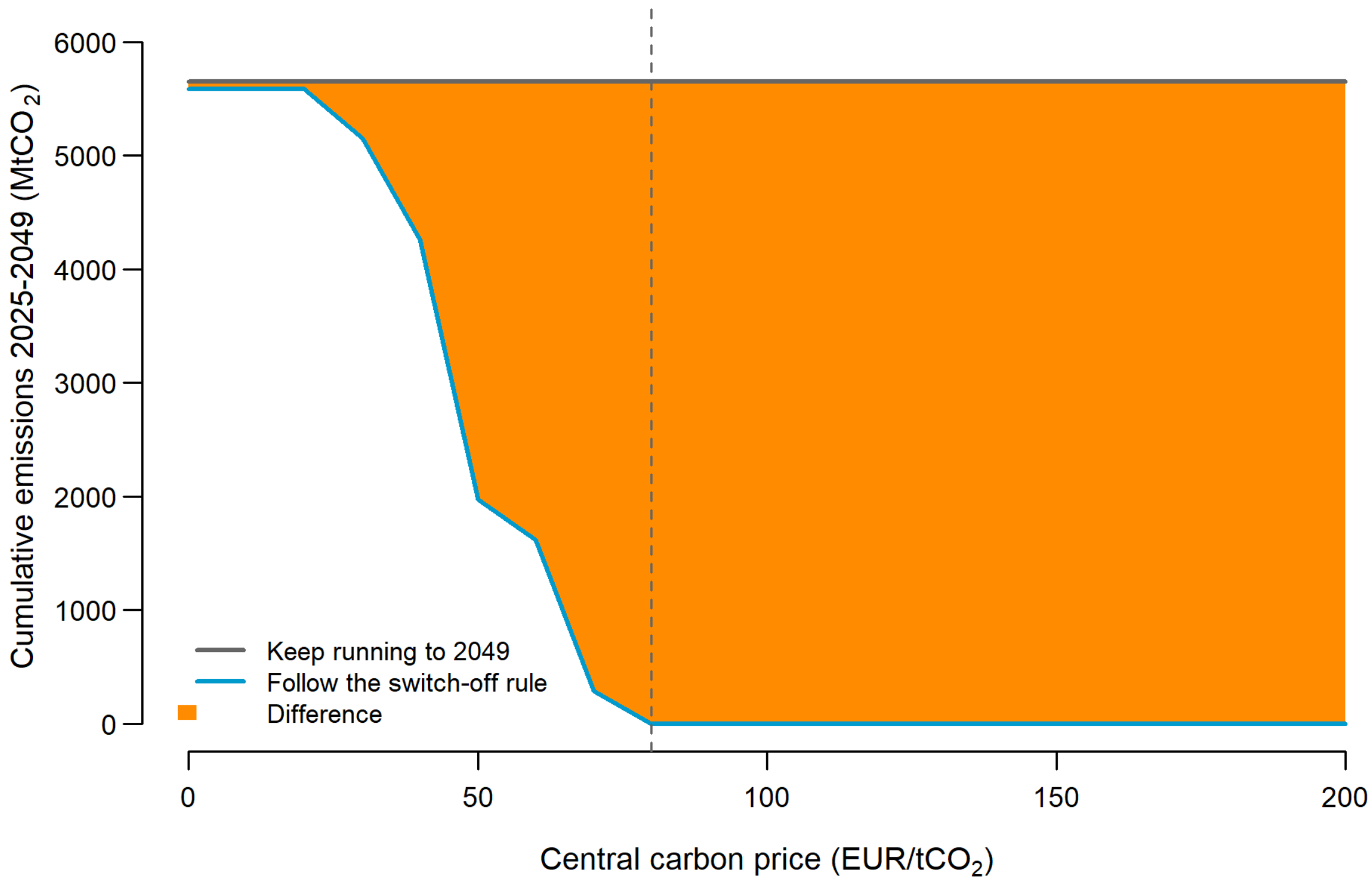

*Figure 6. Cumulative emissions 2025–2049 under both paths, against the central carbon price. The shaded area is the difference. The upper path assumes continued operation of all units and does not account for announced closure dates. Source: own illustration.*

Comparable studies report the consequences of continued operation as foregone plant value rather than as emissions [10], whereas the stranded-asset literature quantifies emissions committed by existing capital rather than emissions arising from an operator's decision rule [4, 5]. Accordingly, the quantity reported here is not directly comparable to either.

## 3.4 Robustness

Because the MDP used in this study contains many static assumptions, the main parameters are tested through a set of sensitivity analyses. This approach determines which parameters matter, which are not decisive for the model's outcomes, and what implications can be drawn from the results.

Figure 7 illustrates the set of parameters tested for their impact on the outcomes, differentiated by power plant type. This is essential because these parameters are static rather than dynamically modeled. The parameters tested are the fuel cost multiplier, electricity price, capacity factor multiplier, discount rate, and the width of the carbon price range ($\sigma$). In all tests, oil plants are not economically viable at any of these values, which is why the oil graph is not affected.

Among the parameters, fuel cost, electricity price, and capacity factor drive the results the strongest. The fuel cost multiplier shows fuel cost ranges of 50% to 150% based on current fuel prices. Lignite is barely affected, hard coal moderately, and natural gas power plants are highly affected. This heterogeneity is due to the dominant cost structure of the different plants. If fuel prices for these plants increase, the plants will shut down at a much lower threshold carbon price. If fuel costs for natural gas are low, e.g., in non-crisis times and with secure supply, the plants are highly profitable and keep operating. Varying the electricity price from 50 to 110 Euro/MWh affects all plants except oil, which are small peakers not very sensitive to the electricity price. The threshold carbon prices for hard coal, lignite, and, most strongly, for natural gas, all increase with higher electricity prices, as it becomes much more profitable to keep operating under higher revenues. This means nearly a tenfold effect for hard coal. At prices of 50-65 Euro/MWh, natural gas is not viable at any carbon price, but at electricity prices of 95-110 Euro/MWh it keeps operating. This shows that the electricity price has, in this model, the largest influence on the threshold carbon price needed to phase out most power plants. As the third most important parameter, the capacity factor multiplier shows what happens if the factor is halved or increased by 150%. The strongest effects are shown for natural gas and hard coal. This is more relevant for low capacity factors, but ultimately the threshold carbon price is not highly affected for high-capacity plants. Low utilization of a plant means that fixed costs dominate and the plant closes almost regardless of carbon prices. Emissions rise and fall almost exactly in proportion to the assumed capacity factor, as halving it gives 2,826 Mt $CO_2$ and increasing it by half gives 8,477 Mt $CO_2$, against 5,651 Mt $CO_2$ emissions at the assumed value.

At the assumed 5% discount rate, the analysis shows that varying it from 3% to 8% does not significantly affect the threshold carbon price, as the curves remain flat. This underscores that the decision to keep operating or to shut down is essentially not influenced by the discount rate.

The width of the carbon price range raises the thresholds steadily, by roughly 3 Euro/t$CO_2$ per 0.25-unit increase in width for hard coal and natural gas, and 3.5 Euro/t$CO_2$ for lignite. This is a relatively uniform response, given that the fuel costs of these plants differ by a factor of twenty. Expected switch-off capacity and emissions at a central carbon price of 80 Euro/t$CO_2$, however, do not change at all: even at the widest range tested, every threshold remains below 80, so the whole fleet still closes in the first period. These columns are therefore identical across all four widths, reflecting where the central price sits relative to the thresholds rather than insensitivity to uncertainty.

Further, setting the plants' annual degradation to zero raises the threshold carbon prices and increases aggregated emissions from operating plants, as expected, because older plants can still produce at the initial output rate and do not face productivity aging. To make closing worthwhile, a higher carbon price would be needed. This is more critical for hard coal plants, as their threshold is affected more strongly because many of these plants are older, compared to natural gas plants, which are barely affected due to being less old.

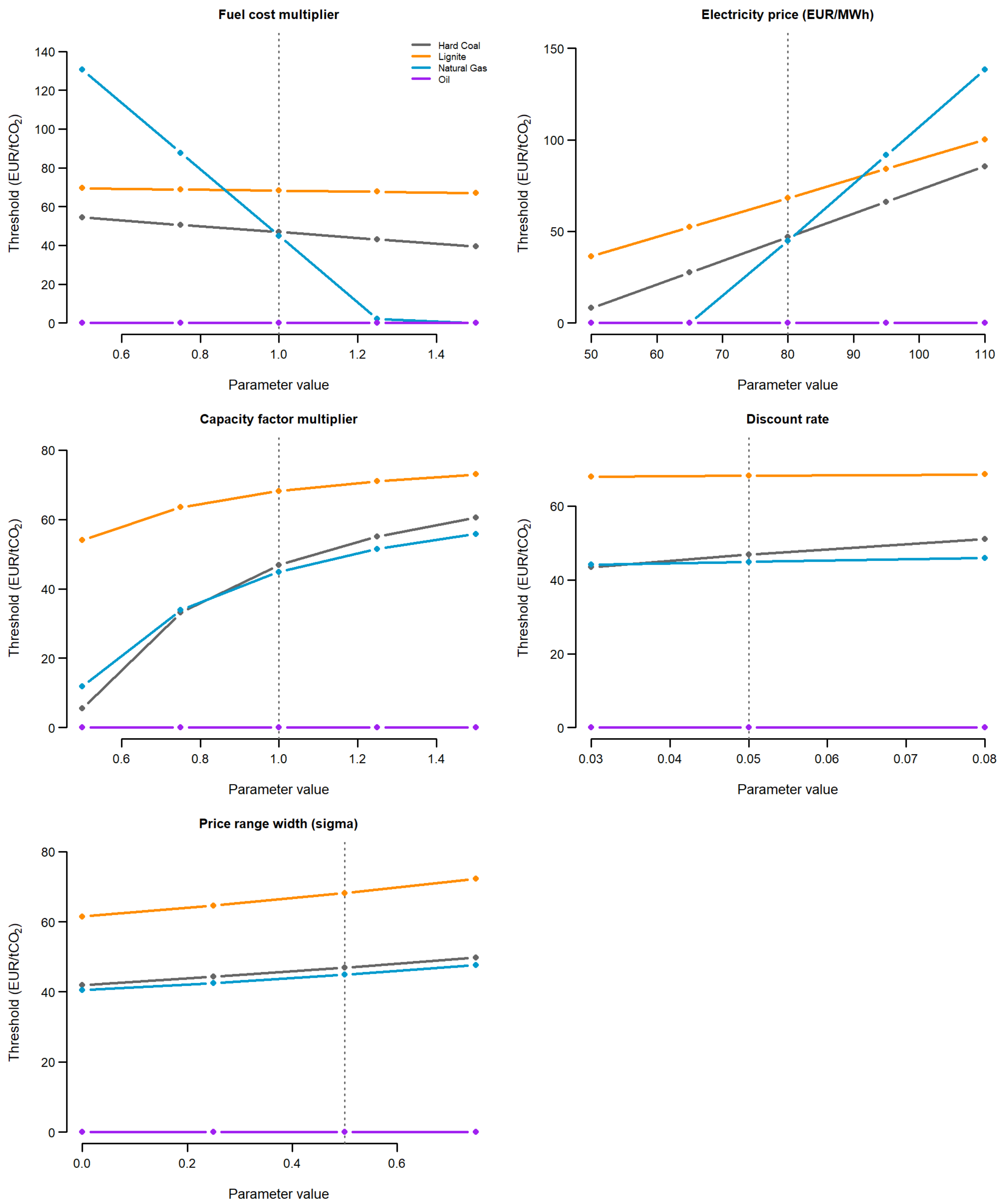

*Figure 7. Switch-off thresholds under variation of each parameter, by fuel. One parameter is varied per panel with the others held at baseline, marked by the dotted line. Panels share the vertical quantity but not its scale. Source: own illustration.*

## 3.5 Limitations

It needs to be stressed that the results of this model are not a forecast, because there is no electricity price response included. The largest omission is that exit at this scale would slow further exit. The fleet curve is therefore much steeper than a general equilibrium model would suggest. Furthermore, there is no constraint on security of supply, no capacity mechanisms, and no country-specific policies like mandated fossil fuel phase-outs, e.g., in Germany's lignite sector. Ignoring the announced closures in the model means that the value of delay is an upper bound here. Additionally, the technology parameters remain illustrative and uniform within a fuel, so the results can only be interpreted with regard to ordering and direction, and do not account for any heterogeneity within a fuel, only between fuels. Accordingly, the sensitivity check of the electricity price shows that assuming static values matters, because varying it alone can shift a fuel's threshold tenfold. This importance does not hold for every parameter to the same extent, but this case illustrates it strikingly. Another issue is that the carbon price and the $\sigma$ matrix are not estimated in detail from EUA price history. In particular, the $\sigma$ matrix captures the uncertainty influencing the plants' decisions, and its size is a result of the persistence of Q and a direct function of $\sigma$ itself.

Still, the assumptions only govern the magnitude of the effects. The qualitative result does not depend on the specific numbers, because any process in which the price might fall back will produce a positive option value. This is not about the exact level of the carbon price, as the model does not assess what the "correct" carbon price is; rather, it is about the shape of the price path. That the uncertainty premium is about 11% also reflects the assumed parametrization, not necessarily the exact size of the premium. Furthermore, operators respond to a price process they treat as exogenous, and there is no stabilizing force from policy action, or the cost of such action, included. The results are therefore not a policy evaluation. E.g., results for coal stations facing carbon costs are more likely to shift to peaking operation than to close, which the fixed capacity factors assumed here do not represent; this reinforces the interpretation of the keep-running path as an upper bound [10].

# 4 Conclusion and policy implications

This study analyzes a set of 850 fossil fuel power plants to examine their optimal phasing-out under a model of repeated decisions by plant operators, and, as a consequence, determines the value of a delayed phase-out if the resulting decision rule is not followed. Under the applied parametrization, the carbon prices at which closing becomes the better choice are 46.9 Euro/tCO2 for hard coal, 68.2 Euro/tCO2 for lignite, and 44.9 Euro/tCO2 for natural gas. This means that half the fleet's capacity is already past its individually optimal phase-out at a carbon price of 37.5 Euro/tCO2, and 99% of the fleet is past the point at 67.5 Euro/tCO2. For the extra carbon price needed because closing is permanent and prices might fall back, continuing to operate the fleet instead of following the phase-out rule would produce 5651 Mt CO2 (at a central carbon price of 80 Euro/tCO2), with the gap widening most steeply in the range of 40 to 70 Euro/tCO2. Oil power plants are excluded here, as they are not viable on energy margins alone at any carbon price.

What is more striking is that uncertainty about the carbon price regime raises the optimal switch-off carbon price threshold by about 11% at a symmetrical $\sigma$ of 0.5 (carbon price range variability around its average). As a consequence, operators keep running their plants even at carbon prices where the model would prescribe closing if they had certainty, because the price process is specified so that the expected carbon price is identical in both cases. The uncertainty allows a decision about continuing or ceasing to operate at a later point under different prices, since a high price may fall back and closure is irreversible. This is what the option value of delay is about. Even under restrictive and static assumptions, the sign, the mechanism, and the near-linear response to how widely the carbon price varies around its average hold and are true irrespective of that.

A stable carbon price closes the same capacity as a volatile one that is approximately a tenth higher; therefore, fossil power plants close earlier if the price is credibly stable. Long-term targets, floor prices, and avoiding sudden reversals have measurable value and will retire capacity early, making a case for reliable, credible carbon price policy. This is, most of all, a statement about the shape of the price path at a constant average, not about a specific carbon price, as the model contains no cost of, e.g., providing stability.

The price will retire the cleanest capacity first. Natural gas-fueled power plants close at a carbon price of 44.9 Euro/tCO2, hard coal at 46.9 Euro/tCO2, and lignite at 68.2 Euro/tCO2. The order is nearly the reverse of carbon intensity, with natural gas emitting the least CO2 and lignite the most. This is because fuel cost, not emissions, determines how much margin a carbon price has to erode. Natural gas plants pay 55 Euro/MWh for fuel and have little to no buffer, while lignite plants' fuel costs are 2.3 Euro/MWh, which can absorb a far higher carbon cost before closing the plant becomes attractive. Consequently, lignite power plants account for 41% of fleet emissions on 15% of capacity; natural gas accounts for 37% of emissions on 55% of capacity. Therefore, the capacity that leaves first is the capacity emitting the least per unit of output. As a result, a uniform carbon price is an imprecise instrument for reducing emissions. Reaching the most carbon-intensive plants requires a price so high that much cleaner capacity has already been removed from the market. In the interim, that cleaner capacity has to be replaced by something else. This implies a transitional period in which the fleet is smaller but more carbon-intensive per unit of output than before. Whether this matters depends on the capacity replacing natural gas power plants. This is currently not represented in the model. However, the ordering depends on relative fuel prices, and robustness shows that the carbon price that makes closing a plant attractive moves from 130.5 Euro/tCO2 to 0 across the fuel cost range. If natural gas is cheap, the order could reverse. What remains is that carbon intensity alone does not determine the exit order, because fuel cost and utilization also determine how large a plant's profitability margin is.

Every carbon price at which closing becomes the better choice lies below 70 Euro/tCO2. The current carbon price, as a result of the EU ETS, is already above the range in which this model predicts closure decisions; therefore, if plants are still running, the carbon price is not what is keeping them open. However, this model also omits several factors that could explain this situation, namely capacity payments, must-run and security-of-supply obligations, network constraints, national policy, and very high electricity prices above this model's assumed values. Which of those impacts the keep-running decision cannot be determined in this study and is left for future work. Still, at current price levels, further increases in the carbon price are unlikely to be the binding instrument for retiring this fleet. That aside, the result that the whole fleet closes at 80 Euro/tCO2 reflects this price relative to the plants' prices at which closing would be profitable; it is not a prediction of imminent plant closures.

The fact that power market conditions can override the carbon price shows that the carbon and electricity prices can work against each other. Varying the electricity price from 50 to 110 Euro/MWh shifts the price at which retiring, e.g., a hard coal power plant, would be preferable from about 8 to 85 Euro/tCO2. Natural gas plants are unviable at electricity prices of 50-65 Euro/MWh but survive well enough at 95-110 Euro/MWh. A carbon price calibrated to retire capacity under one set of power market conditions will not work under another set of circumstances, and the relevant conditions, e.g., gas crises, demand shocks, extreme weather, are outside carbon policy's control. Consequently, the effectiveness of the carbon price is conditional on circumstances it cannot govern, making a case for monitoring the combined signal rather than the carbon price alone. It is not an argument that carbon pricing fails.

The carbon price at which oil power plants are better off retiring is zero, demonstrating that a plant's survival can be determined entirely by revenue that the carbon price does not affect, as plants operate under very high fixed costs with little capacity. These oil plants are small peakers that earn from scarcity pricing and capacity payments, neither of which the model includes. For plants like that, carbon pricing essentially does not influence the closure decision. Wherever capacity mechanisms or scarcity rents are material, this situation would recur. This is consistent with evidence from a German lignite plant, whose continued operation was found to depend on whether subsidies were included in the valuation rather than on its energy margins [12]. As fossil power plants across the fleet run fewer hours, they would earn less from energy and more from availability, which progressively reduces the carbon price's leverage over them. Therefore, carbon pricing and capacity remuneration can pull in opposite directions. If capacity payments keep plants available that would otherwise be unviable under energy margins alone, the two instruments need to be designed with reference to each other. However, this is a structural observation, not a recommendation about any particular capacity market, as this study's model does not include one.

Several extensions would change these results, and they are not equally consequential. The one most likely to change the model outcome is the electricity price feedback. The second most likely impact would be caused by using a calibrated carbon process, and the third impact would be capacity market revenue. If capacity payments are introduced, the order in which the fuels' thresholds are reached could be reversed, and oil plants are not viable on their own but would be when compensated in situations of supply shortage. A fourth factor affecting the outcome would be plant-level capacity factors, as they would allow for heterogeneity within fuels. Lastly, the mandated plant closure schedule would also affect the outcomes.

Still, what this model delivers is not a date for Europe's fossil phase-out but a way to interpret the carbon price. Closing a power plant is irreversible, so what matters to an operator is not only how high the price is but also how firmly it is expected to remain at that level. A credible carbon price retires fossil fuel power plants based on their own profitability incentives much earlier than a higher but more uncertain price.

# Acknowledgments

During the preparation of this work, the author used DeepL and Claude AI to improve readability and language. After using these tools, the author reviewed and edited the manuscript as needed and takes full responsibility for the content of the publication.

# Funding

This work was supported in part by the Helmholtz Association under the program "Energy System Design". Open Access is funded by the Deutsche Forschungsgemeinschaft (DFG, German Research Foundation) (491111487).

# Availability of data and materials

Data will be made available on request.

# Declaration of interests

The authors declare that they have no known competing financial interests or personal relationships that could have appeared to influence the work reported in this paper.

# References

1. Tong, D., Q. Zhang, Y. Zheng, K. Caldeira, C. Shearer, C. Hong, Y. Qin, and S.J. Davis. *Committed emissions from existing energy infrastructure jeopardize 1.5 °C climate*

*target*. Nature, 2019. **572**(7769): p. 373-377, https://doi.org/10.1038/s41586-019-1364-3.
2. Fofrich, R., D. Tong, K. Calvin, H.S. De Boer, J. Emmerling, O. Fricko, S. Fujimori, G. Luderer, J. Rogelj, and S.J. Davis. *Early retirement of power plants in climate mitigation scenarios*. Environmental Research Letters, 2020. **15**(9): p. 094064, https://doi.org/10.1088/1748-9326/ab96d3.
3. Fofrich Navarro, R., L. Liebermann, F.C. Moore, C. Shearer, and S.J. Davis. *Ownership of power plants stranded by climate mitigation*. Nature Sustainability, 2026. **9**(2): p. 328-336, https://doi.org/10.1038/s41893-025-01707-5.
4. Pfeiffer, A., R. Millar, C. Hepburn, and E. Beinhocker. *The '2°C capital stock' for electricity generation: Committed cumulative carbon emissions from the electricity generation sector and the transition to a green economy*. Applied Energy, 2016. **179**: p. 1395-1408, https://doi.org/10.1016/j.apenergy.2016.02.093.
5. Mercure, J.F., H. Pollitt, J.E. Viñuales, N.R. Edwards, P.B. Holden, U. Chewpreecha, P. Salas, I. Sognnaes, A. Lam, and F. Knobloch. *Macroeconomic impact of stranded fossil fuel assets*. Nature Climate Change, 2018. **8**(7): p. 588-593, https://doi.org/10.1038/s41558-018-0182-1.
6. Abrell, J. *EUETS.INFO*. 2026. https://www.euets.info/.
7. Dixit, A.K. and R.S. Pindyck. *Investment under Uncertainty*. 1994: Princeton University Press. ISBN: null. http://www.jstor.org/stable/j.ctt7sncv. DOI: https://doi.org/10.2307/j.ctt7sncv.
8. McDonald, R. and D. Siegel. *The Value of Waiting to Invest*. The Quarterly Journal of Economics, 1986. **101**(4): p. 707-727, https://doi.org/10.2307/1884175.
9. Pindyck, R. *Irreversibility, Uncertainty, and Investment*. National Bureau of Economic Research Working Paper Series, 1990, http://dx.doi.org/10.3386/w3307.
10. Abadie, L.M., J.M. Chamorro, and M. González-Eguino. *Optimal Abandonment of EU Coal-fired Stations*. The Energy Journal, 2011. **32**(3): p. 175-207, https://doi.org/10.5547/ISSN0195-6574-EJ-Vol32-No3-7.
11. Fleten, S.-E. and E. Näsäkkälä. *Gas-fired power plants: Investment timing, operating flexibility and CO2 capture*. Energy Economics, 2010. **32**(4): p. 805-816, https://doi.org/10.1016/j.eneco.2009.08.003.
12. Glensk, B. and R. Madlener. *Evaluating the enhanced flexibility of lignite-fired power plants: A real options analysis*. Energy Conversion and Management, 2018. **177**: p. 737-749, https://doi.org/10.1016/j.enconman.2018.09.062.
13. Rust, J. *Optimal Replacement of GMC Bus Engines: An Empirical Model of Harold Zurcher*. Econometrica, 1987. **55**(5): p. 999-1033, https://doi.org/10.2307/1911259.
14. Medury, A. and S. Madanat. *Incorporating network considerations into pavement management systems: A case for approximate dynamic programming*. Transportation Research Part C: Emerging Technologies, 2013. **33**: p. 134-150, https://doi.org/10.1016/j.trc.2013.03.003.
15. Chao, W., Y. Shi, Q. He, J. Chen, and K. Zheng. *Multi-critic and transformer-enhanced optimization dispatch for energy management systems*. Sustainable Energy Technologies and Assessments, 2025. **84**: p. 104734, https://doi.org/10.1016/j.seta.2025.104734.
16. Wang, X. and H. Zhang. *Valuation of CCS investment in China's coal-fired power plants based on a compound real options model*. Greenhouse Gases: Science and Technology, 2018. **8**(5): p. 978-988, https://doi.org/10.1002/ghg.1809.

17. Liu, H., Q. Wang, Y. Chi, L. Xu, Y. Luo, J. Ye, and N. Zhou. *Multi-stage coordinated dynamic reactive power planning for wind-penetrated power systems considering coal-fired power plant retirement*. International Journal of Electrical Power & Energy Systems, 2026. **181**: p. 112100, https://doi.org/10.1016/j.ijepes.2026.112100.
18. Koch, N., S. Fuss, G. Grosjean, and O. Edenhofer. *Causes of the EU ETS price drop: Recession, CDM, renewable policies or a bit of everything?—New evidence*. Energy Policy, 2014. **73**: p. 676-685, https://doi.org/10.1016/j.enpol.2014.06.024.
19. Blyth, W., R. Bradley, D. Bunn, C. Clarke, T. Wilson, and M. Yang. *Investment risks under uncertain climate change policy.* Energy Policy, 2007. **35**(11): p. 5766-5773, https://doi.org/10.1016/j.enpol.2007.05.030.
20. Puterman, M.L. John Wiley & Sons, Inc. *Markov Decision Processes: Discrete Stochastic Dynamic Programming*. Wiley Series in Probability and Statistics, 1994. https://doi.org/10.1002/9780470316887.
21. Chadès, I., G. Chapron, M.-J. Cros, F. Garcia, and R. Sabbadin. *MDPtoolbox: a multi-platform toolbox to solve stochastic dynamic programming problems*. Ecography, 2014. **37**(9): p. 916-920, https://doi.org/10.1111/ecog.00888.
22. R Core Team. R Foundation for Statistical Computing. *R: A language and environment for statistical computing*. 2026. https://www.R-project.org/.
23. Gotzens, F., H. Heinrichs, J. Hörsch, and F. Hofmann. *Performing energy modelling exercises in a transparent way - The issue of data quality in power plant databases*. Energy Strategy Reviews, 2019. **23**: p. 1-12, https://doi.org/10.1016/j.esr.2018.11.004.
24. Eurostat;. *Electricity prices for non-household consumers - bi-annual data*. 2025. https://doi.org/10.2908/NRG_PC_205.
25. European Commission. *Trends in carbon intensity and the macroeconomic role of the EU Emissions Trading System*. 2025. https://economy-finance.ec.europa.eu/trends-carbon-intensity-and-macroeconomic-role-eu-emissions-trading-system_en.
26. Rogelj, J., G. Luderer, R.C. Pietzcker, E. Kriegler, M. Schaeffer, V. Krey, and K. Riahi. *Energy system transformations for limiting end-of-century warming to below 1.5 °C*. Nature Climate Change, 2015. **5**(6): p. 519-527, https://doi.org/10.1038/nclimate2572.
27. Quaschning, V. *Spezifische Kohlen-dioxid-emissionen verschiedener Brennstoffe*. 2026. https://www.volker-quaschning.de/datserv/CO2-spez/index.php.
28. Jesse, B.-J., S. Morgenthaler, B. Gillessen, S. Burges, and W. Kuckshinrichs. *Potential for Optimization in European Power Plant Fleet Operation*. Energies, 2020. **13**, 718, https://doi.org/10.3390/en13030718.
29. Ember. *Electricity Data Explorer*. 2026. https://ember-energy.org/data/electricity-data-explorer/.
30. Lorenczik, S., J.H. Keppler, P. Fraser, S. Bilbao y Leon, L. Mir, G. Sousa, S. Kim, B. Wanner, and C. Donovan. International Energy Agency (IEA), Nuclear Energy Agency (NEA). *Projected Costs of Generating Electricity*. 2020. https://www.iea.org/reports/projected-costs-of-generating-electricity-2020.
31. S&P Global Platts. *Turning tides - The future of oil after IMO 2020*. 2019. https://www.spglobal.com/platts/en/market-insights/special-reports/oil/fuel-oil-after-imo-2020.

32. Ea Energy Analyses A/S. Danish Energy Agency. *Balmorel Lite Assumptions*. 2014. https://www.balmorellite.dk/assumptions.php.
33. Kost, C., P. Müller, J. Sepulveda Schweiger, V. Fluri, and J. Thomsen. Fraunhofer Institute for Solar Energy Systems ISE. *Levelized cost of electricity - Renewable energy technologies*. 2024. https://www.ise.fraunhofer.de/content/dam/ise/en/documents/publications/studies/EN2024_ISE_Study_Levelized_Cost_of_Electricity_Renewable_Energy_Technologies.pdf.
34. Jindal, A. and G. Shrimali. *Cost–benefit analysis of coal plant repurposing in developing countries: A case study of India*. Energy Policy, 2022. **164**: p. 112911, https://doi.org/10.1016/j.enpol.2022.112911.
35. Raimi, D. Resources for the Future (RFF). *Decommissioning US Power Plants: Decisions, Costs, and Key Issues*. 2017. https://www.rff.org/documents/578/RFF20Rpt20Decommissioning20Power20Plants.pdf.
36. Climate Council. *New research reveals our ageing coal fleet is unreliable and risks energy security*. 2025. https://www.climatecouncil.org.au/resources/new-research-reveals-our-ageing-coal-fleet-is-unreliable-and-risks-energy-security/.

# Appendix

| Symbol | Meaning |
|---|---|
| $\mathrm{cap}_i$ | Nameplate capacity, MW |
| $\Lambda_f$ | Net retirement payment, $\Lambda_f = V_f - D_f$, Euro/MW |
| $\Omega_f(\sigma)$ | Option value of waiting, $Omega_f(\sigma) = \tau_f(\sigma) - \tau_f(0)$, Euro/tCO2 |
| $D_f$ | Decommissioning cost, Euro/MW |
| $E^{\mathrm{open}}$ | Cumulative emissions 2025-2049 under continued operation, MtCO2 |
| $E^{\mathrm{rule}}$ | Cumulative emissions 2025-2049 under the optimal policy, MtCO2 |
| $R_f(s, a)$ | Reward per MW of capacity in one period, Euro |
| $V_f$ | Salvage value, Euro/MW |
| $W_t(s)$ | Value function per MW |
| $e_{it}$ | Annual emissions of plant i in period t, MtCO2 |
| $g_f(\alpha)$ | Share of nameplate output retained at age a, declining with plant age |
| $k_i$ | Age bin of plant i in 2025 |
| $m_f$ | Fixed operating and maintenance cost, Euro/MW/year |
| $\bar{p}$ | Central carbon price, Euro/t$CO_2$ |
| $p^c(c)$ | Carbon price in state c, Euro/t$CO_2$ |
| $p^e$ | Electricity price, Euro/MWh |
| $q_f(k)$ | Output of a plant of one MW in one period, MWh |
| $\alpha_k$ | Midpoint age of bin k, years |
| $\delta_f$ | Annual degradation rate |
| $\varepsilon_f$ | Carbon intensity, tCO2/MWh |
| $\eta_{it}$ | Probability that plant i is still operating in period t |
| $\kappa_f$ | Capacity factor, share of the year at full output |
| $\pi_t(s)$ | Optimal policy, the best action in state s in period t |
| $\tau_f(\sigma)$ | Switch-off threshold at width $\sigma$: the lowest central carbon price at which switching off immediately is optimal, Euro/tCO2 |
| $\phi_f$ | Fuel cost, Euro/MWh |
| $\Delta$ | Years per period, 5 |
| $\Delta E$ | $E^{\mathrm{open}} - E^{\mathrm{rule}}$, MtCO2 |
| $H$ | Hours per period, 8760 $\Delta$ |
| $P(s' \mid s, a)$ | Probability of moving to state $s'$ from s under action a |
| $Q$ | Carbon price transition matrix, $Q_{cc'} = \Pr(c_{t+1} = c' \mid c_t = c)$ |
| $T$ | Number of decision periods, 5 |
| $a$ | Action, $a = 1$ keep running, $a = 2$ switch off |
| $c,\ c'$ | Carbon price state in the current and following period, $c \in \{\mathrm{L}, \mathrm{M}, \mathrm{H}\}$ |
| $f$ | Fuel: hard coal, lignite, natural gas, oil |
| $f(i)$ | Fuel of plant i |
| $i$ | Plant, $i = 1, \dots, I$ |
| $k$ | Age bin, k$= 1, \dots, K; K = 12$ |
| $r$ | Annual discount rate |
| $s,\ s'$ | Model state, $s = (c, k)$, or the absorbing retired state; $\mathcal{S}$ is the set of all states |
| $t$ | Decision period, $t = 1, \dots, T\$(2025{,}2030, \dots{,}2045)$ |
| $\beta$ | Per-period discount factor, $(1 + r)^{-\Delta}$ |
| $\sigma$ | Width of the carbon price range, as a share of the central price |

*Table 6. Variable description.*

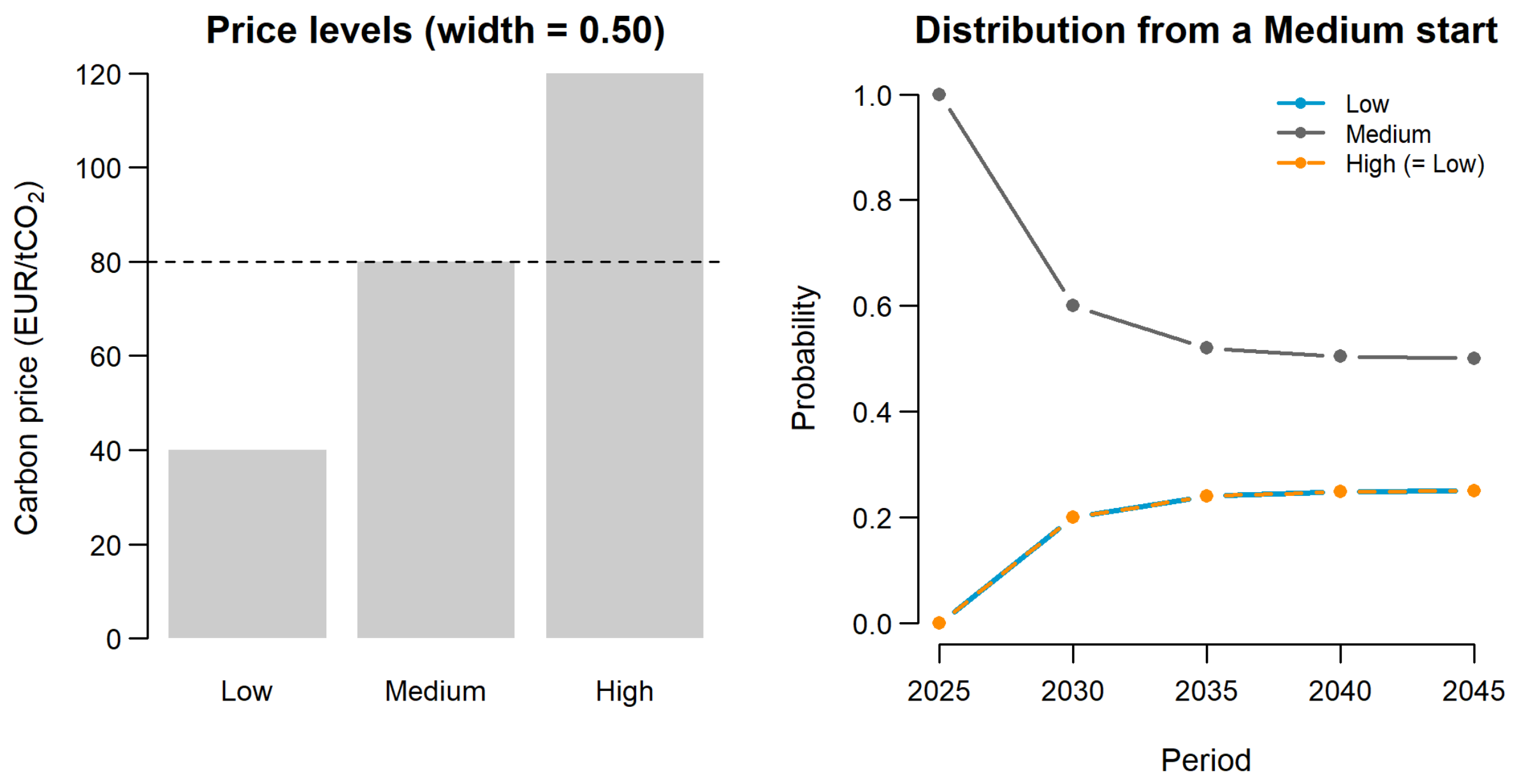

*Figure 8. The modeled carbon price process. The left panel shows the three price levels at the width used in the main results, placed symmetrically around the central price. The right panel shows how the probability of occupying each level evolves over the five decision periods from a start in the* medium *state. The* low *and* high *lines coincide exactly, because the transition matrix is unchanged when the order of the states is reversed.* High *is drawn dashed so that both remain visible. This symmetry is what holds the expected carbon price equal to the central price in every period and at every width, and is the property of the comparison in Section 3.2.*

| From | Low | Medium | High | LongRun |
|---|---|---|---|---|
| Low | 0.55 | 0.4 | 0.05 | 0.25 |
| Medium | 0.2 | 0.6 | 0.2 | 0.5 |
| High | 0.05 | 0.4 | 0.55 | 0.25 |

*Table 7. Carbon price transition matrix and its long-run distribution. Rows give the current state and columns the state in the following period; entries are probabilities and each row sums to one. The matrix is a modeling choice rather than an estimate from allowance price data. It is unchanged when the order of the states is reversed, which is the property that keeps the expected carbon price constant across all widths tested.*

| Fuel | AgeBin (years) | Threshold (Euro/tCO2) | Static (Euro/tCO2) |
|---|---|---|---|
| Hard Coal | 55+ | 46.9 | 53.3 |
| Lignite | 40-44 | 68.2 | 62.6 |
| Natural Gas | 15-19 | 44.9 | 39.3 |
| Oil | 55+ | 0 | -84.5 |

*Table 8. Switch-off thresholds and static breakevens for the representative plant of each fuel, at the carbon price range width used in the main results. The threshold is the lowest central carbon price at which switching off immediately is optimal, located by bisection to a tolerance of 0.1 Euro/tCO2. A threshold of zero indicates that the plant is not viable on energy margins at any carbon price, and a negative static breakeven has no economic interpretation; both apply to oil.*

| Test | Value | Hard Coal | Lignite | Natural Gas | Oil |
|---|---|---|---|---|---|
| Fuel cost multiplier | 0.5 | 54.4 | 69.5 | 130.5 | 0 |
| Fuel cost multiplier | 0.75 | 50.6 | 68.8 | 87.7 | 0 |
| Fuel cost multiplier | 1 | 46.9 | 68.2 | 44.9 | 0 |
| Fuel cost multiplier | 1.25 | 43.1 | 67.7 | 2.1 | 0 |
| Fuel cost multiplier | 1.5 | 39.4 | 67 | 0 | 0 |
| Electricity price (Euro/MWh) | 50 | 8.2 | 36.4 | 0 | 0 |
| Electricity price (Euro/MWh) | 65 | 27.6 | 52.4 | 0 | 0 |
| Electricity price (Euro/MWh) | 80 | 46.9 | 68.2 | 44.9 | 0 |
| Electricity price (Euro/MWh) | 95 | 66.2 | 84.1 | 91.6 | 0 |
| Electricity price (Euro/MWh) | 110 | 85.5 | 100.1 | 138.2 | 0 |
| Capacity factor multiplier | 0.5 | 5.5 | 54.1 | 11.9 | 0 |
| Capacity factor multiplier | 0.75 | 33.1 | 63.5 | 33.9 | 0 |
| Capacity factor multiplier | 1 | 46.9 | 68.2 | 44.9 | 0 |
| Capacity factor multiplier | 1.25 | 55.1 | 71 | 51.5 | 0 |
| Capacity factor multiplier | 1.5 | 60.6 | 73 | 55.8 | 0 |
| Discount rate | 0.03 | 43.5 | 68 | 44.1 | 0 |
| Discount rate | 0.05 | 46.9 | 68.2 | 44.9 | 0 |
| Discount rate | 0.08 | 51.1 | 68.6 | 45.9 | 0 |
| Price range width ($\sigma$) | 0 | 41.9 | 61.5 | 40.5 | 0 |
| Price range width ($\sigma$) | 0.25 | 44.3 | 64.6 | 42.5 | 0 |
| Price range width ($\sigma$) | 0.5 | 46.9 | 68.2 | 44.9 | 0 |
| Price range width ($\sigma$) | 0.75 | 49.8 | 72.2 | 47.6 | 0 |

*Table 9. Switch-off thresholds by fuel under each parameter variation. Each test varies one parameter and holds the remainder at baseline. The fuel cost and capacity factor tests scale each fuel's own value rather than assigning a common one, so the columns show each fuel's sensitivity to its own parameter. Expected switch-off capacity in the same runs is evaluated at a central carbon price of 80 Euro/tCO2.*

| Test | Value | KeepRunning (Mt) | FollowRule (Mt) | Difference (Mt) |
|---|---|---|---|---|
| Fuel cost multiplier | 0.5 | 5651 | 1980 | 3671 |
| Fuel cost multiplier | 0.75 | 5651 | 1328 | 4323 |
| Fuel cost multiplier | 1 | 5651 | 0 | 5651 |
| Fuel cost multiplier | 1.25 | 5651 | 0 | 5651 |
| Fuel cost multiplier | 1.5 | 5651 | 0 | 5651 |
| Electricity price (Euro/MWh) | 50 | 5651 | 0 | 5651 |
| Electricity price (Euro/MWh) | 65 | 5651 | 0 | 5651 |
| Electricity price (Euro/MWh) | 80 | 5651 | 0 | 5651 |
| Electricity price (Euro/MWh) | 95 | 5651 | 2793 | 2859 |
| Electricity price (Euro/MWh) | 110 | 5651 | 4490 | 1161 |
| Capacity factor multiplier | 0.5 | 2826 | 0 | 2826 |
| Capacity factor multiplier | 0.75 | 4238 | 0 | 4238 |
| Capacity factor multiplier | 1 | 5651 | 0 | 5651 |
| Capacity factor multiplier | 1.25 | 7064 | 0 | 7064 |
| Capacity factor multiplier | 1.5 | 8477 | 0 | 8477 |
| Discount rate | 0.03 | 5651 | 0 | 5651 |
| Discount rate | 0.05 | 5651 | 0 | 5651 |
| Discount rate | 0.08 | 5651 | 0 | 5651 |
| Price range width ($\sigma$) | 0 | 5651 | 0 | 5651 |
| Price range width ($\sigma$) | 0.25 | 5651 | 0 | 5651 |
| Price range width ($\sigma$) | 0.5 | 5651 | 0 | 5651 |
| Price range width ($\sigma$) | 0.75 | 5651 | 0 | 5651 |

*Table 10. Cumulative emissions 2025-2049 under both paths for the same parameter variations as Table 9, evaluated at a central carbon price of 80 Euro/tCO2. Identical values across the rows of a test indicate that the whole fleet switches off in the first period regardless of that parameter, which reflects the position of the central carbon price relative to the thresholds rather than insensitivity to the parameter.*